\documentclass[useAMS,usenatbib]{mn2e}
\usepackage{epsfig}

\def\der{\rm d}
\def\Mc{M_{\rm i}}
\def\tc{t_{\rm i}}
\def\Mi{M_{\rm f}}
\def\Ri{R_{\rm f}}
\def\ti{t_{\rm f}}
\def\rs{r_{\rm s}}
\def\ra{r_{\rm a}}
\def\roc{\rho_{\rm c}}
\def\rou{\rho_{\rm u}}
\def\rocr{\rho_{\rm crit}}
\def\delc{\delta_{\rm c}}
\def\delm{\Delta_{\rm m}}
\def\PS{_{\mathrm t}}
\def\nbody{{$N$}-body }

\title[Halo Structure]{On the Origin of the
Internal Structure of Haloes}

\author[Manrique, Salvador-Sol\'e \& Raig]{Alberto
Manrique, Eduard Salvador-Sol\'e and Andreu Raig\\
Departament d'Astronomia i Meteorologia, Universitat de
Barcelona, Mart{\'\i} Franqu\`es 1, E-08028 Barcelona, Spain\\
E-mail: Alberto.Manrique@am.ub.es (AM), eduard@am.ub.es (ESS), 
araig@am.ub.es (AR)}

\begin{document}

\maketitle

\begin{abstract}

High-resolution $N$-body simulations of hierarchical cosmologies have
shown that the density and velocity dispersion profiles of dark-matter
haloes display well-definite universal forms whose origin remains
unknown. In the present paper, we calculate the internal structure of
haloes expected to arise in any such cosmologies by simply taking into
account that halo growth proceeds through an alternate sequence of
discrete major mergers and long periods of gentle accretion. Major
mergers cause the violent relaxation of the system subject to the
boundary conditions imposed by accreting layers beginning to fall in
at that moment. Accretion makes the system develop inside-out from the
previous seed according to the spherical infall model. The predicted
structure is in very good agreement with the results of numerical
simulations, particularly for moderate and low mass haloes. We find
strong indications that the slight departure observed in more massive
systems is not due to a poorer theoretical prediction, but to the more
marked effects of the limited resolution used in the simulations on
the empirical profiles. This may have important consequences on the
reported universality of halo structure.

\end{abstract}

\begin{keywords}
cosmology -- dark matter -- galaxies: formation, evolution
\end{keywords}

\section{INTRODUCTION}\label{intro}
The dominant dark component of matter in the universe appears to be
clustered in bound haloes which form the skeleton of all astronomical
objects of cosmological interest, from dwarf galaxies to rich galaxy
clusters. All we know about these systems comes from their
gravitational effects on the luminous matter they trap or on the light
traveling across them. This is so little information that the
following fundamental questions have prevailed for a long time. Does
the internal structure of haloes depend on their mass? Does it depend
on their past history? And on cosmology? How is this structure set?

Gunn \& Gott (1972; \citealt{gott75}; \citealt{gunn77}) were the first
to address this problem by considering the simplified case of haloes
forming through spherical infall, that is, the collapse of a density
fluctuation of smooth, isotropically distributed, dissipationless
matter in an otherwise homogeneous expanding universe. Under the
adiabatic invariant collapse approximation, they derived the density
profile arising from some specific initial conditions. The effects of
refining the derivation used and of adopting more and more realistic
initial conditions have been subsequently addressed in a series of
works (\citealt{fg84}; \citealt{berts85}; \citealt{hs85};
\citealt{rg87}; \citealt{ryd88}; \citealt{zh93}; \citealt{lok00};
\citealt{LH00}; \citealt{pgrs}; \citealt{ekp}; \citealt{nuss}).

An important result of this research line is that haloes grow
inside-out as new material is being incorporated in secondary
infall. That is, despite the continuous shell-crossing (and radial
relaxation, see below) of the infalling layers with the previously
relaxed body, the structure remains at any moment essentially
unaltered within the instantaneous radius. As shown below, this sole
consideration should permit us to determine the halo density profile
provided the infall rate of matter is known. Much progress has been
achieved in the last twenty years in the modelling of halo growth. The
extended Press-Schechter (PS) model (\citealt{PS}; \citealt{b};
\citealt{BCEK}; \citealt{LC93}) makes quite accurate predictions,
indeed, on the rate at which haloes increase their mass \citep{LC94}
in hierarchical cosmologies like the one describing the real
universe. Unfortunately, haloes develop in such cosmologies through
continuous mergers rather than through smooth spherical infall.

The effects of mergers depend on the relative mass of the
progenitors. Major mergers bring the whole system out of equilibrium
making it move in the phase space around some attractor. During this
process, particles experience random accelerations owing to the
rapidly varying collective potential well which causes the relaxation
of the system \citep{LB67} in a similar although more dramatic way
than in the case of spherically infalling layers. This violent
relaxation erases any imprint of the previous history of the system,
the new equilibrium state reached being characterised by a normal
distribution of particle velocities similar to that yielded by the
common two-body relaxation, although independent of particle
mass. Were haloes isolated and, hence, unperturbed after any such
dramatic event they would end up as infinite, spherical systems with a
uniform, monomass, isotropic velocity dispersion. This is the reason
why haloes are often modelled as isotropic, isothermal spheres (e.g.,
\citealt{King72}; \citealt{SIR99}). But haloes are not isolated
systems. During the violent relaxation process, they keep on
collecting matter through minor mergers making the boundary conditions
to vary in some unknown way. This severely limits the predictive power
of the violent relaxation theory (e.g., \citealt{s87}).

To gain a deeper insight on the internal structure of haloes many
authors have turned to numerical experiments. Using high-resolution
cosmological $N$-body simulations Navarro, Frenk \& White (1996,
1997) have found that the spherically averaged density profile of
haloes of all masses is well fitted by the simple analytical
expression
\begin{equation}
\rho(r)={\roc \rs^3 \over r(\rs + r)^2} 
\label{nfw}
\end{equation} 
in any hierarchical cosmology analysed. In equation (\ref{nfw}), $r$
is the radial distance to the halo centre, and $\rs$ and $\roc$ are
the halo scale radius and characteristic density, respectively. The
latter two parameters are related to each other and to the mass $M$ of
the halo through the condition that the virial radius $R$ of the
system encloses, by definition, an average density equal to a fixed
factor $f$ times some density $\rou$ characterising the universe at
that moment. The values of $f$ found in the literature are in the
range [178, 500], while $\rou$ is taken equal to $\rocr$, the critical
density for closure, or $\bar\rho$, the mean cosmic density. Navarro
and collaborators adopted $f=200$ and $\rou=\rocr$.

On the other hand, the spherically averaged, locally isotropised,
velocity dispersion profile $\Sigma(r)$ is found to be well fitted by
the solution of the Jeans equation for hydrostatic equilibrium and
negligible rotation,
\begin{equation}
\Sigma^2(r)\left( {\der \ln \Sigma^2 \over \der \ln r} + 
{\der \ln \rho \over \der \ln r} \right) = 
- {3\,G M(r) \over r}
\label{heq}
\end{equation}
with $\rho(r)$ the function given in equation (\ref{nfw}) and $M(r)$
the corresponding mass within $r$, by adopting the boundary condition
of null pressure at infinity. Note that, although the hydrostatic
condition is natural to hold, the boundary condition at infinity is
not obvious owing to the limited extent of haloes.

These empirical results have been confirmed by other authors
(\citealt{CL97}; \citealt{HJS99a}; \citealt{Bull01}), there being some
controversy only at very small radii (\citealt{Moore98};
\citealt{JS00}; \citealt{FM01}) where they are the most affected by
the spatial resolution of the simulations. In any event, the
fundamental question about the origin of that empirical halo structure
remains. It is not even clear whether this is a general result of
gravitational collapse, including smooth spherical infall, or the
specific consequence of repeated mergers (\citealt{HJS99b};
\citealt{SW98}; \citealt{Moore99}; \citealt{LH00}).

In the present paper, we use a variant of the extended PS model
distinguishing between minor and major mergers (\S\ \ref{merg}) to
derive the structure of haloes predicted in hierarchical cosmologies
and compare it with that found in numerical simulations (\S\
\ref{theo}). The result of this comparison supports the idea that the
internal structure of haloes is the natural imprint of their
hierarchical growth (\S\ \ref{diss}).

\section{MINOR AND MAJOR MERGERS}\label{merg}

The scaling of the previous density profile varies with cosmology
although the same trend is always found: the characteristic density
$\roc$ decreases with increasing halo mass. \citet{NFW97} suggested
that this universal trend is due to the fact that, in hierarchical
cosmologies, more massive haloes form later when the mean density of
the universe is lower. Unfortunately, the idea that haloes fix their
structural properties at formation is hard to assess because the own
concepts of halo formation and destruction are rather fuzzy in
hierarchical cosmologies in which haloes grow through continuous
mergers.

To properly define these concepts it is convenient to distinguish
between minor and major mergers (\citealt{ms96}; \citealt{ks96};
\citealt{ssm}; \citealt{pmp00}; \citealt{cbw01}). In the Modified
Press-Schechter (MPS) model (\citealt{ssm}; \citealt{rgs98};
\citealt{rgs01}) mergers producing a fractional mass increase above a
given threshold $\delm\sim 0.5$ are regarded as major, and minor
otherwise. Note that a given merger may be seen as major or minor
depending on the viewpoint of the partner halo which is
considered. Since haloes are essentially unperturbed in minor mergers
while they are completely rearranged in major ones we say that haloes
{\it survive\/} in the former case and {\it are destroyed\/} in the
latter. The destruction of a halo does not necessarily imply the
formation of a new one: the largest partner partaking of the merger
can see it as minor and, hence, survive to it being identified to the
final system. This kind of mergers therefore correspond to the {\it
accretion\/} by the most massive halo of the remaining partners. (The
definitions of minor and major mergers given above imply that there is
at most one surviving halo in any given merger.) Only those mergers in
which {\it all\/} partners are destroyed or, equivalently, no merging
halo survives, do mark the {\it formation\/} of new haloes.

The MPS model yields analytical expressions for the mass accretion
rate of haloes (see eq. [\ref{accr}] below), their destruction and
formation rates, as well as the distribution probability functions of
progenitor masses and of formation and destruction times whose median
values define, respectively, the typical mass of progenitors and the
typical halo formation and destruction times. All these theoretical
quantities are in good agreement with the results of \nbody
simulations \citep{rgs01}.

In particular, the average rate at which mass is accreted onto a halo
with $M$ at $t$ is, according to the definition of $\delm$, 
\begin{equation}
r_{\rm a}(M,t)=\int_M^{M(1+\delm)}\,(M'-M)
\,r\PS(M, M',t)\,\der M'\, , \label{accr}
\end{equation}
where
\begin{eqnarray}
r\PS(M, M',t)\;\der M' =
{\sqrt{2/\pi}
\over\sigma^2(M')}
{\der\delc\over \der t}\,{\der\sigma(M')\over\der M'} 
\left[1-{\sigma^2(M')\over\sigma^2(M)}\right]^{-3/2}\label{umr}
\nonumber \\ 
\times\exp\left\{-{\delc^2(t)\over 
2\sigma^2(M')}\left
[1-{\sigma^2(M')\over\sigma^2(M)}\right]\right\}\,\der M'\qquad
\end{eqnarray}
is the instantaneous transition rate at $t$ from haloes with $M$ to
haloes between $M'$ and $M'+\der M'$ due to mergers of {\it whatever
amplitude} provided by the original extended PS model \citep{LC93}. In
equation (\ref{umr}), $\delc(t)$ is the linear extrapolation to the
present time $t_0$ of the critical overdensity of primordial
fluctuations collapsing at $t$, and $\sigma(M)\equiv \sigma(M,t_0)$ is
the r.m.s. fluctuation of the density field at $t_0$ smoothed over
spheres of mass $M$; both $\delc(t)$ and $\sigma(M)$ depend on
cosmology.

The $M(t)$ track followed during accretion by haloes with a given mass
$\Mc$ at a given time $\tc$ is therefore the solution of the
differential equation
\begin{equation}
{\der M\over\der t}=r_{\rm a}[M(t),t] \label{act}
\end{equation}
for the initial conditions $M(\tc)=\Mc$. Strictly speaking, this
solution is the {\it average} track followed by those haloes. Real
accretion tracks actually diffuse from it owing to the effects of
individual random minor mergers, the scatter remaining nonetheless
quite limited along the typical lifetime of haloes (see
\citealt{rgs01}).

As shown by \citet{rgs01}, the formation of haloes corresponds to rare
binary mergers between similarly massive progenitors. Hence, they do
cause the rearrangement of the corresponding global systems, the
resulting violent relaxation yielding a more or less spherical haloes
with new density profiles independent of their past history. In
contrast, minor mergers are frequent multiple events, particularly in
the case of very small captured haloes. The graininess of accreted
matter can therefore be neglected in a first approximation, and the
resulting configuration (i.e., a central more or less spherical,
relaxed object surrounded by a rather smooth distribution of matter
falling into it) can be well approximated by the spherical infall
model. Consequently, haloes should grow inside-out during the
accretion phase.

\section{Theory vs. simulations}\label{theo}

The point of view that haloes grow through long periods of gentle
accretion after their violent formation in major mergers is clearly
much more accurate than the two extreme points of view so far
considered in order to derive analytical density profiles (see \S\
\ref{intro}), namely, the cases of pure spherical infall or one
single isolated major merger. During the accretion phase, the
inside-out growth condition holds and, since the average mass
accretion rate is well-known (eq. \ref{accr}), one can readily infer
the structure developing around the initial seed. On the other hand,
the structure of this latter, fixed at formation, should also be
possible to determine since the boundary conditions met by the system
during violent relaxation are well-known, too: they are set by
the accretion regime starting at that moment.

\subsection{Density profile}

Suppose a spherical halo formed at $\ti$ with mass $\Mi$ and radius
$\Ri$. The typical $M(t)$ track followed by the halo during the
subsequent accretion phase is therefore the solution of the
differential equation (\ref{act}) with initial condition
$M(\ti)=\Mi$. According to the spherical infall model, the accreted
mass is deposited, at any moment $t$, at the instantaneous radius $R(t)$
without altering the inner density profile. Consequently, we have
\begin{equation}
M(t)-\Mi=\int_{\Ri}^{R(t)} 4 \pi r^2 \rho(r)\,\der r
\end{equation} 
with $\rho(r)$ the developing density profile at $t$. By
differentiating this relation taking into account the abovementioned
definition of the instantaneous radius
\begin{equation}
R(t)=\left[ {3\,M(t) \over f 4 \pi \rou(t)} \right]^{1/3} 
\label{rad}
\end{equation} 
and equation (\ref{act}) we are led to the expression
\begin{equation}
\rho(t)= f \rou \left\{ 1-{M(t) \over \ra[M(t),t]}
{\der \ln \rou \over \der t} \right\}^{-1}    
\label{dens}
\end{equation}
for the density at $r=R(t)$. Equations (\ref{rad}) and (\ref{dens})
therefore define, in the parametric form, the wanted density profile
at any radius $r\geq\Ri$.

But this is not enough to explain the shape of the density profile
drawn from simulations. Indeed, a significant fraction of it (of order
50 \%) corresponds to the inner region, $r < \Ri$, fixed at formation
(see the position of $\Ri$ in Fig. 2). To derive the density profile
in that inner region we shall take into account that the structure
emerging from violent relaxation at $\ti$ must satisfy the boundary
conditions imposed by accretion starting at that moment. This implies
that the inner and outer profiles of any physical quantity must match
up at $\Ri$. Some discontinuity would only appear in case that these
boundary conditions were incompatible with the inner structure. But,
violent relaxation takes some time to proceed during which surrounding
layers begin to be accreted. Hence, the final steady state, far from
being incompatible with those boundary conditions, will perfectly adapt
to them.

Both $M(t)$ and $\rou(t)$ entering in equations (\ref{rad}) and
(\ref{dens}) are known functions which can be differentiated to any
arbitrary order. Consequently, one can derive the values of $\rho(r)$
and any order derivative of it at any radius in the outer region, in
particular at the matching radius $\Ri$. Then, by taking the Taylor
series expansion of $\rho(r)$ at $\Ri$ one can estimate that function
at any radius $r < \Ri$ to any arbitrary accuracy. Although this
proves the existence of one unique inner solution, a much practical
way to derive it is the following one. As mentioned in \S\ \ref{merg},
all haloes having the same mass at a given time follow the same
accretion track along the common time interval since their respective
formation times $\ti$, and since the functions $M(t)$ and $\rou(t)$
coincide for all these haloes in that common time interval, the same
is true for the respective outer profiles in the corresponding radial
ranges. Moreover, the inner solutions of all these haloes must
coincide with the outer solutions of those having undergone the last
major merger early enough since the profiles and derivatives of any
order of these latter take identical values at the radii $\Ri$ of the
former and, therefore, have identical Taylor series expansions at such
radii. Strictly speaking, this conclusion refers to the {\it average}
density profile since individual profiles will show some scatter
around it coming from the abovementioned scatter in real $M(t)$
tracks.

The previous reasoning proves from the mathematical point of view that
all haloes with a given mass at a given time have the same (average)
density profile regardless of their specific formation time, and that
this unique density profile, which naturally appears after the past
history of the system has been erased, coincides with the outer
solution of haloes formed at an arbitrarily early epoch. From the
physical point of view the reason for this apparently surprising
result is well understood. As well known, in the case of spherical
infall there is only one mass distribution at any moment preserving
the steady state of the inner system despite the shell-crossing of the
infalling layers; this corresponds to the solution of the spherical
infall model for a given accretion rate. In the case of violent
relaxation subsequent to a major merger, the resulting steady
structure must also adapt to the shell crossing of the infalling
layers beginning to be accreted at that moment. This therefore
necessarily leads to the same inner mass distribution as in the case
of spherical infall at that moment. Note that this explains the result
obtained by \citet{HJS99b} that the density profile of haloes closely
resembles the NFW profile regardless of whether they form in a
hierarchical cosmology or via smooth spherical collapse.

The exact values of $f$ and $\rou$ used to delimit relaxed haloes in
simulations or in the real universe are a mere convention and do not
obviously affect their shape. However, from equations (\ref{rad}) and
(\ref{dens}) we see that the density profile here predicted depends on
the definition of halo radius adopted. There is no contradiction
however between these two facts. That definition is crucial, in
general, for the ability of the PS model to correctly describe the
halo growth process. For instance, the Press-Schechter mass function
only fits the empirical mass function provided some specific
definition of halo radius as this fixes the mass of objects. Likewise,
we cannot pretend to recover the actual density profile of haloes from
the PS formalism for any arbitrary definition of halo radius; this
must be chosen so to obtain the best agreement between the predictions
of the PS model and the results of simulations. Note however that such
an ``accreting radius'' must not necessarily coincide with the
arbitrary ``delimiting radius'' used to compare the resulting
theoretical profile with the empirical one obtained by any specific
author. In this latter case, we will operate with the theoretical
profile previously derived just as if it were the real profile of a
simulated (or observed) halo: we shall apply the same delimiting
radius which may or may not coincide with the accreting radius
previously used. This can be readily done in most cases without any
further consideration. Only in the case that the value of $f$ used in
the derivation of the profile is larger than the one used to delimit
it in the final comparison, or that $\rocr$ is used in the derivation
while $\bar\rho$ is used in the comparison some extension of the
theoretical profile towards the future will be required. But this
simply reflects the real situation met in dealing with simulated (or
observed) haloes: their actual steady region extends well beyond any
popular delimiting radius.

\begin{figure}
\centerline{
\psfig{figure=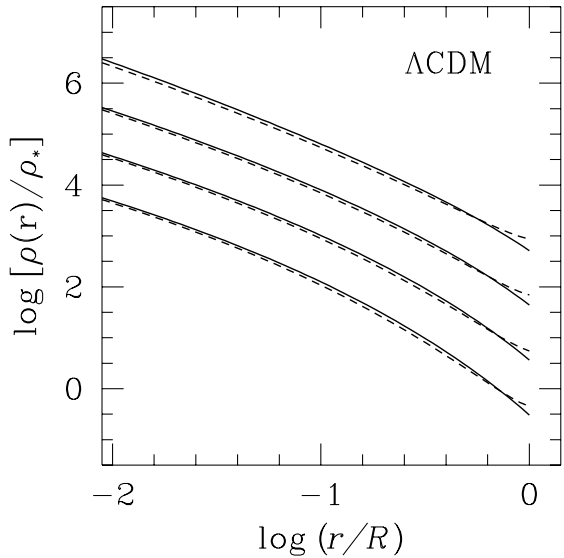,width=8 truecm}}
\vskip -.2cm
\caption{Theoretical density profiles for current haloes of masses
10$^{-2} M_\ast$, 10$^{-1} M_\ast$, $M_\ast$, and 10 $M_\ast$ (from bottom
to top) in a flat, $\Omega_{\rm m}=0.25$, CDM model inferred using
$\rou=\rocr$ (dashed lines) and $\rou=\bar\rho$ (full lines). In the
plot we use the same delimitation of haloes as in Navarro et
al. (1996, 1997). $M_\ast$ is the mass scale at which primordial
density fluctuations leave the linear regime, and $\rho_\ast$ stands
for $\rocr M_\ast/M$.}
\bigskip
\end{figure}

The correct behaviour of the extended PS model has only been checked
in detail in the case of scale-free cosmologies (\citealt{LC94};
\citealt{rgs01} for the MPS version). It is therefore necessary to
establish the best value of $\rou$ to be used in case that $\rocr$ and
$\bar\rho$ do not coincide. In Figure 1, we plot the theoretical
profiles inferred from $\rou=\rocr$ and $\rou=\bar\rho$ in a flat,
$\Omega_{\rm m}=0.25$ CDM cosmology. As can be seen, $\rou=\bar\rho$
leads to a decreasing density profile with monotonous varying slope
while $\rou=\rocr$ leads to a density profile which tends to level off
at large radii. (The radii where the two solutions deviate from each
other correspond to cosmic times at which $\Lambda$ becomes
non-negligible.) Since such a behaviour is not observed in simulated
(or observed) haloes, we conclude that the best value of $\rou$ to use
is $\bar\rho$. Concerning the value of $f$ we have checked that the
theoretical density profile is always very insensitive to it, any
value in the range 178--500 yielding fully indistinguishable results
at the resolution of Figure 1. Hereafter, we adopt $f=200$.

\begin{figure*}
\psfig{figure=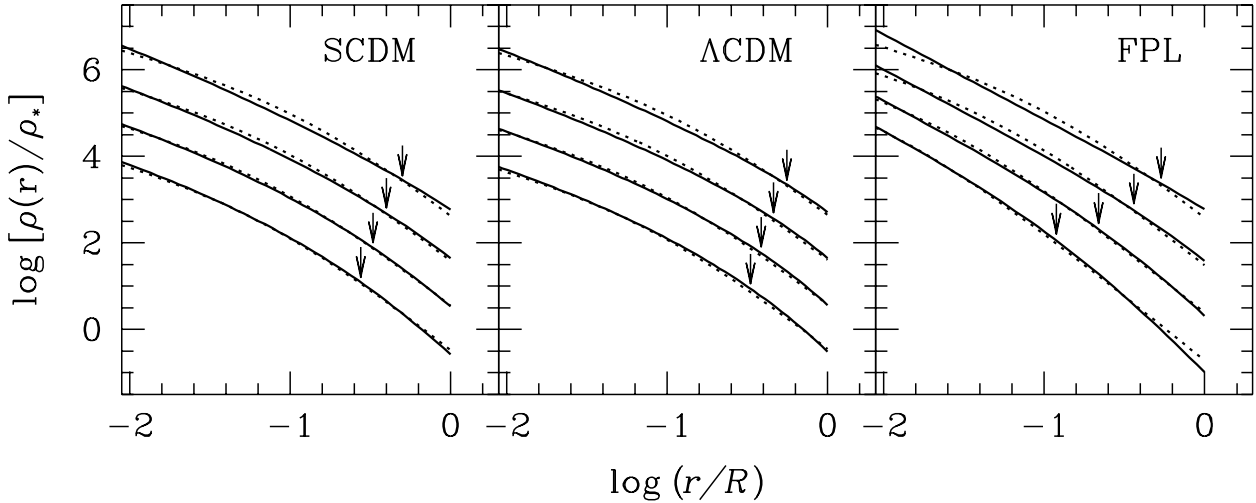,width=17 truecm}
\vskip -10cm
\caption{Predicted (full lines) and empirical (dotted lines) density
profiles for current haloes with the same masses as in Fig. 1 in the
three following cosmologies analysed by \citet{NFW97}: a flat,
$\Lambda=0$, CDM model (SCDM), a flat, $\Omega_{\rm m}=0.25$, CDM
model ($\Lambda$CDM), and a flat, $\Lambda=0$, model with power-law
spectrum of density fluctuations of index $n=-1$ (FPL).  Arrows mark
the frontier between the inner regions fixed at the {\it typical\/}
formation time and outer ones developed during the subsequent
accretion phase.}
\end{figure*}

\begin{figure*}
\centerline{
\psfig{figure=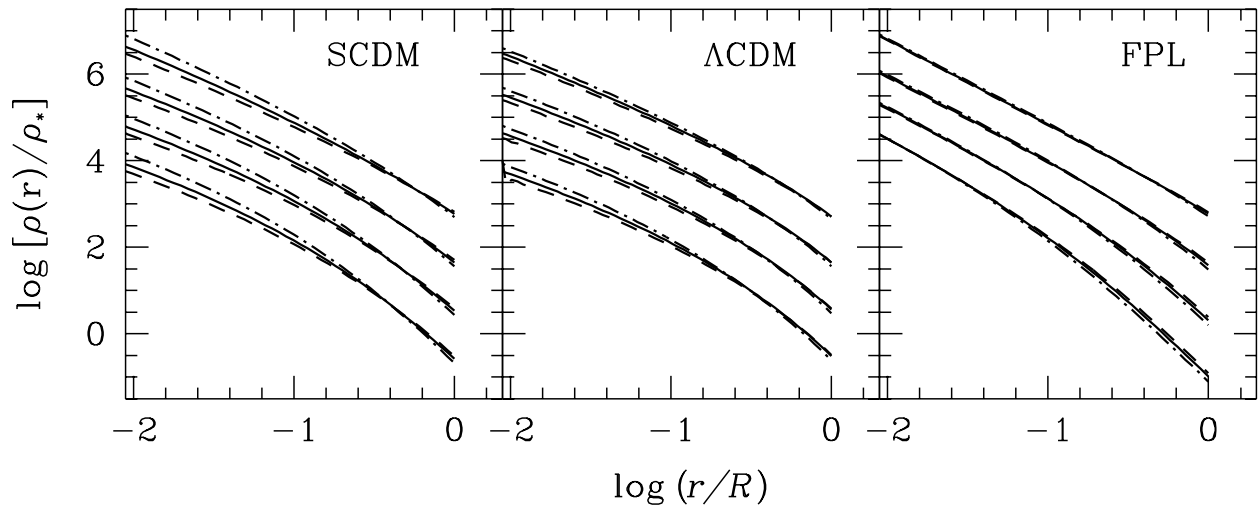,width=17 truecm}}
\vskip -10cm
\caption{Same density profiles as in Fig. 2 drawn using a value of
the threshold for major mergers, $\delm$, equal to 0.5 (full lines)
compared to the profiles resulting from $\delm=0.3$ (dot-dashed lines)
and $\delm=0.7$ (dashed lines).}
\bigskip
\end{figure*}

In Figure 2, we compare the average density profiles predicted for
current haloes of four different masses in three distinct cosmologies
with the corresponding empirical profiles obtained by
\citet{NFW97}. As can be seen, there is good agreement between theory
and simulations, particularly for moderate and low mass haloes. A
similar result is found in any cosmology analysed. We want to stress
that, once we have fixed the value of the effective threshold $\delm$
for major mergers entering in the MPS model, the predicted density
profiles of haloes of any mass at any time are completely determined
both in shape and normalization. Thus, there is no freedom to play
with. As mentioned in \citet{rgs01}, the comparison between theory and
simulations in terms of the description of halo growth alone does not
allow us to constrain the value of $\delm$. The only restriction is
given a priori by the expected dynamical effects of mergers of
different amplitude which points to the plausible range [0.3, 0.7]. In
Figure 3 we show the effects of adopting any of these two extreme
values of $\delm$. By comparing these results with those obtained from
the most reasonable central value of $0.5$ shown in Figure 2 we
conclude that the value of $\delm$ is, on the contrary,
well-constrained around that central value by the internal structure
of haloes.

\begin{figure*}
\centerline{
\psfig{figure=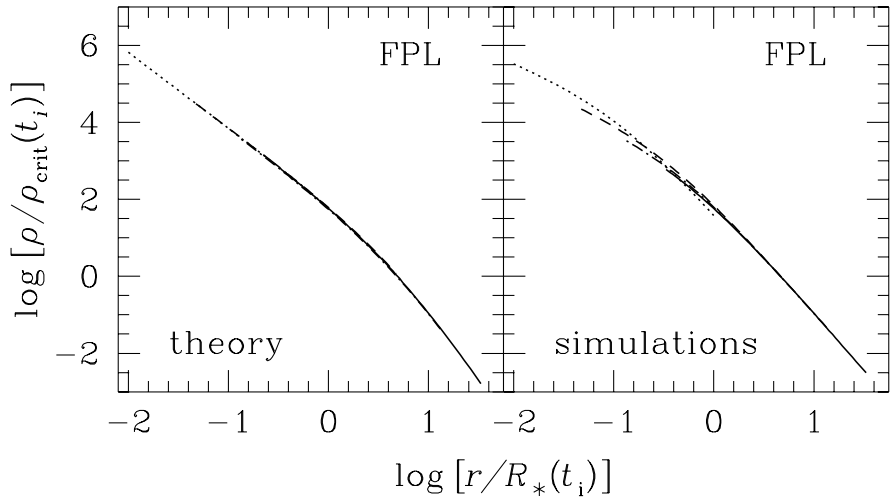,width=12 truecm}}
\vskip -5.2cm
\caption{Same density profiles as in the right panel of Fig. 2 but
conveniently shifted so to apply to haloes of the four different
masses $\Mc$ but at four appropriate times $\tc$ making them
indistinguishable in that scale-free cosmology (provided the
assumption of inside-out growth during accretion). For clarity we have
separated the theoretical profiles derived here (left panel) from the
empirical ones obtained by \citet{NFW97} identically shifted (right
panel). The shifts applied to the profiles at $t_0$ can be
reconstructed from the different symbols used for the non-overlapping
empirical profiles: 10$^{-2} M_\ast$ (full line), 10$^{-1} M_\ast$
(dot-dashed line), $M_\ast$ (dashed line), and 10 $M_\ast$ (dotted
line). $R_\ast(t)$ is the scale length corresponding to $M_\ast(t)$.}
\end{figure*}

The observed trend for the agreement between theory and simulations to
degrade towards large halo masses cannot be due to a bad modelling of
the halo growth by means of the MPS model. As previously mentioned,
the correct behaviour of the mass accretion rate has been checked
against numerical simulations in scale-free cosmologies \citep{rgs01}
such as the one plotted in the right panel of Figure 2 where the
departure is the most apparent. To understand the possible origin of
this effect we will concentrate on this particular case. In scale-free
cosmologies two haloes with different masses, $M$ and $M'$, taken at
$t$ and $t'$ satisfying the condition $\sigma(M,t)=\sigma(M',t')$ are
indistinguishable. All the density profiles shown in Figure 2
correspond to the present time $t_0$, but the inside-out growth
condition guarantees that, at any other epoch, haloes have the same
profiles though truncated at different radii enclosing the
corresponding masses. Thus, by conveniently rescaling the density
profiles of haloes of different masses at $t_0$, they must perfectly
overlap. In Figure 4 we show the result of applying such shifts on
both the four theoretical density profiles shown in the right panel of
Figure 2 and the corresponding empirical profiles obtained by
\citet{NFW97}. (Shifted density profiles are only displayed down to
the minimum initial radii where simulated data were fit.) As can be
seen, the theoretical density profiles satisfy the expected homologous
condition while the empirical profiles do not. There are two possible
reasons for this latter fact: either the finite resolution of the
simulations introduces some artificial scale breaking the theoretical
homologous behaviour of density profiles, or haloes significantly
deviate, during accretion, from the inside-out growth implied by the
adiabatic invariant collapse approximation. In the former case, one
should trust on the theoretical profiles derived here, and conversely
in the latter. The fact that the deviation of the empirical profiles
from the homologous condition only concerns very small radii where the
inside-out growth condition during accretion should only play an
indirect role and where the density profiles are the most affected by
the finite resolution of the simulations strongly suggests that there
may be some problem, indeed, with the empirical profiles there. It
would therefore be of great interest to check through numerical
simulations of scale-free cosmologies whether or not large mass haloes
grow inside-out during accretion as expected from the adiabatic
invariant collapse approximation.

The previous discussion is extremely relevant in connection with the
problem on the actual slopes of density profiles at very small radii
($r \sim 10^{-2}\,R$). In general, we find good agreement with the
slopes reported by \citet{NFW97} (see Fig. 2). But, for large masses
in the scale-free case, we find a much steeper slope (as large as $-2$
in logarithmic units). We want to remark that this is necessary for
those density profiles to satisfy the abovementioned homologous
condition since the slope at small radii of haloes of a given mass
must match the slope at larger radii of more massive haloes. (Note
that the asymptotic slope of $-2$ in the scale-free case is
independent, indeed, of the value of $\delm$; see Fig. 3.) If the
correct behaviour of the theoretical profiles derived here is
confirmed we will be led to the conclusion that the density profile of
very massive haloes deviates from the form reported by Navarro and
collaborators, and that the asymptotic slope at small radii of the
halo density profile actually depends on the particular cosmology
considered.

\subsection{Velocity dispersion profile}

And what about the velocity dispersion profile $\Sigma(r)$? This
profile could be readily derived from the Jeans equation (\ref{heq})
using the previous density profile provided that the value of $\Sigma$
were known at some radius. But it is not. (We are avoiding of course
making any unjustified assumption reporting to infinity.) Since, from
the dynamical point of view, the unique density profile derived above
could perfectly exist side by side with an infinite variety of
velocity dispersion profiles the following questions rises. Why does
the empirical velocity dispersion profile of haloes show one unique
form? What fixes it? As mentioned, in the case that the system has
undergone just one unique violent relaxation far in the past and no
subsequent accretion of matter we would expect it with a {\it
uniform\/} velocity dispersion. But this solution is incompatible with
the density profile previously calculated imposed by accretion. This
suggests the following natural guess: the velocity dispersion profile
emerging from violent relaxation at the formation of a halo should be
as close as possible to uniform, the only small deviation being forced
by the boundary conditions imposed by accretion at that moment.

\begin{figure*}
\centerline{
\psfig{figure=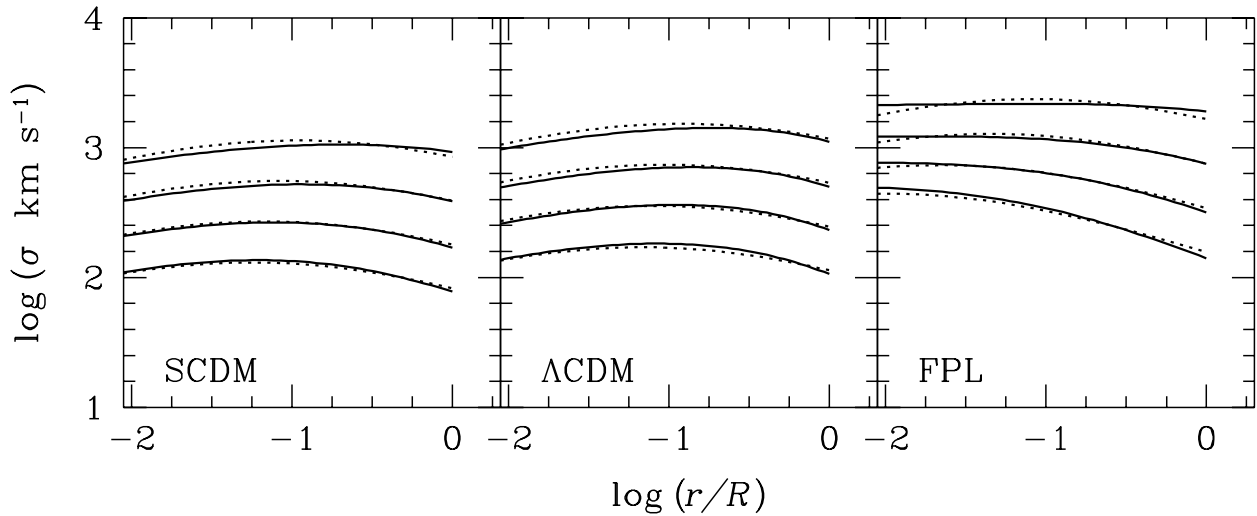,width=17 truecm}}
\vskip -10cm
\caption{Same as Figure 1 but for the velocity dispersion profile.}
\end{figure*}

To derive such a guessed velocity dispersion profile we will follow an
iterative procedure. For the zero-order solution of $\Sigma^2$ at
$r\le\Ri$ we take the square value of the uniform velocity dispersion
that violent relaxation tends to establish, given by equation
(\ref{heq}) with null logarithmic derivative of $\Sigma^2$. Then we
compute the zero-order value of $\Sigma^2$ corresponding to a slightly
larger or smaller, arbitrarily close radius and from the two values we
estimate the zero-order logarithmic derivative of $\Sigma^2$ at $r\le
\Ri$. By substituting this derivative into the Jeans equation
(\ref{heq}) we obtain the more correct first-order value of $\Sigma^2$
at that radius. Finally, by repeating this procedure, we can obtain an
estimate as close as wanted to the true value of $\Sigma^2(r)$ at any
radius of the inner region. This iterative procedure leads to one
unique profile among the infinite number that satisfy the Jeans
equation for the density profile derived above using the mass
accretion rate. Since we have started the iteration with the values of
$\Sigma(r)$ corresponding to an inner uniform velocity dispersion
profile (not satisfying that equation), the final solution is
necessarily the closest to the latter uniform profile which is, at the
same time, compatible with the boundary conditions imposed by
accretion as wanted.

Once the value of $\Sigma^2(\Ri)$ is known we can use the Jeans
equation (\ref{heq}) with the universal density profile derived above
to infer the velocity dispersion profile in the outer region. But we
can also apply the previous iterative procedure directly to any radius
of the outer region. Indeed, the fact that both solutions satisfy the
Jeans equation and that they take the same value at $\Ri$ guarantees
that the two methods yield identical results. Moreover, since haloes
formed at different times with initially relaxed regions overlapping
over some radial range have, by construction, identical velocity
dispersion profiles in the common radial ranges, we are also led to the
conclusion that there is one unique velocity dispersion profile for
all haloes having the same mass at a given time regardless of their
particular formation time, and that this unique velocity dispersion
profile coincides with the inner solution, given by the iterative
procedure above, of haloes formed at an arbitrarily late time.

In Figure 5 we compare, for the same cosmologies as in Figures 2, the
theoretical velocity dispersion profiles, inferred using the previous
iterative procedure in the whole range of radii, with the empirical
profiles obtained by \citet{NFW97}. As can be seen the predictions are
as good as in the case of the density profiles. This confirms
our guess on the origin of the universal velocity dispersion profile.

\section{DISCUSSION}\label{diss}

It is generally believed that the PS formalism describes the mass
growth of dark-matter haloes in hierarchical cosmologies but does not
tell us anything about their internal structure. This is however in
contradistinction with the idea that such a structure arises just from
gravitation (a scale-free force) and {\it the way that dark-matter
haloes grow}. In the present paper, we have shown that, when the
distinct dynamical effects of minor and major mergers are taken into
account, the PS model also makes definite predictions on the internal
structure of haloes, the resulting average theoretical profile being
in very good agreement with that drawn from high-resolution numerical
simulations. Our results therefore prove that the structure of haloes,
including their scaling with mass, is the natural consequence of the
combined action of minor and major mergers: outside the total radius
of the system at formation the structure is fully determined by the
rate at which mass is accreted, while, inside that radius, it results
from the initial violent relaxation with boundary conditions set by
accretion at that moment. The behaviour of the theoretical profiles
derived here strongly suggests that the larger the halo mass, the more
apparent are the effects on the respective density profiles of the
limited resolution of the simulations. The confirmation of this
suspicion would imply a different density profile for very massive
haloes from that reported by Navarro, Frenk \& White (1996, 1997) and,
what is more important, the non-universality of halo structure.

The aim of the present paper was not to derive the accurate
distribution of matter in real haloes, but to understand the origin of
their empirical apparently universal structure. For this reason we
have considered the simplified case of pure dark-matter haloes as
those dealt with in \nbody experiments. In the real universe, about
one tenth of the halo mass is in the dissipative baryonic component,
which might have appreciable effects. Likewise, we have assumed
spherical symmetry and neglected any rotation of haloes as well as any
anisotropy of the velocity tensor while, in hierarchical cosmologies,
haloes have a slight angular momentum and, what is more important,
they are immersed in large filamentary structures making them accrete
matter preferentially along one privileged direction \citep{West94}
and feel, in their final steady configuration, the tidal field of such
anisotropic structures \citep{ss93}. On the other hand, even in the
case of exact spherical symmetry, some velocity anisotropy would
emerge in the outer part of haloes owing to the distinct evolution of
the radial and tangential velocity dispersions in accreted layers
during infall. All these secondary effects explain the elongation and
slight angular momentum and velocity anisotropy observed in real as
well as simulated haloes. (Note however that the density profile
derived here is independent of the actual degree of anisotropy of the
velocity tensor.) Finally, in the present paper, we have focused on
the structure of haloes at $z=0$, the only redshift for which accurate
empirical data for different cosmologies are available
(\citealt{NFW97}). In a forthcoming paper, we will study the predicted
dependence on redshift of this structure and compare it with the
results of numerical simulations carried out by \citet{Bull01} for a
$\Lambda$CDM cosmology.

\vspace{0.75cm} \par\noindent 
{\bf ACKNOWLEDGMENTS} \par 

\noindent This work was supported by the Spanish DGES grant
AYA2000-0951. AM acknowledges the hospitality of the CIDA staff in
M\'erida (Venezuela) where part of this work was carried out.

\end{document}